# Title

# Prediction of transonic flow over supercritical airfoils using geometric-encoding and deep-learning strategies

*Zhiwen Deng[a,*], Jing Wang[b,*], Hongsheng Liu[a], Hairun Xie[c], BoKai Li[a], Miao Zhang[c], Tingmeng Jia[a], Yi Zhang[a], Zidong Wang[a] and Bin Dong[d1]*

[a] *Huawei Technologies Co., Ltd*

*Shenzhen 518129, China*

[b]*School of Aeronautics and Astronautics, Shanghai Jiao Tong University, Shanghai, 200240, China*

[c] *Shanghai Aircraft Design and Research Institute, Shanghai 200436, China,*

[d] *Center for Machine Learning Research, Peking University, Beijing 100871, China*

[*]*These authors contributed equally to this work*

---

[1] Corresponding author:

*E-mail address:* dongbin@math.pku.edu.cn

# Abstract

The Reynolds-averaged Navier–Stokes equation for compressible flow over supercritical airfoils under various flow conditions must be rapidly and accurately solved to shorten design cycles for such airfoils. Although deep-learning methods can effectively predict flow fields, the accuracy of these predictions near sensitive regions and their generalizability to large-scale datasets in engineering applications must be enhanced. In this study, a modified vision transformer-based encoder–decoder network is designed for the prediction of transonic flow over supercritical airfoils. In addition, four methods are designed to encode the geometric input with various information points and the performances of these methods are compared. The statistical results show that these methods generate accurate predictions over the complete flow field, with a mean absolute error on the order of 1e-4. To increase accuracy near the shock area, multilevel wavelet transformation and gradient distribution losses are introduced into the loss function. This results in the maximum error that is typically observed near the shock area decreasing by 50%. Furthermore, the models are pretrained through transfer learning on large-scale datasets and fine-tuned on small datasets to improve their generalizability in engineering applications. The results generated by various pretrained models demonstrate that transfer learning yields a comparable accuracy from a reduced training time.

# 1. Introduction

Supercritical airfoils are critical design components of modern civil aircraft [1] that considerably influence their aerodynamic characteristics, such as cruise efficiency, drag divergence, and stall behaviors. Thus, aerodynamic analysis and optimization during the industrial design of supercritical airfoils involves many iterations of numerical simulations [2]. Reynolds-averaged Navier-Stokes (RANS) modeling [3] is the mainstream tool for such simulations, but it is time-consuming to apply RANS models for repeated iterative calculations. Thus, to shorten the design cycle for supercritical airfoils, there is a need for a method that can rapidly and accurately predict flow fields around such airfoils.

Recently, deep-learning methods have emerged as promising alternatives to RANS models for rapidly predicting flow fields [4] and have attracted significant interest in airfoil research [5–9]. These deep-learning methods are one of two types. The first type are a combination of traditional computational fluid dynamics (CFD) solvers with artificial intelligence (AI) methods and rapidly generate high-quality solutions [4,5,10]. For example, Singh et al. [5] introduced turbulence modeling for forecasting separated flow around airfoils through adjoint-based data assimilation and AI-based feature mapping. They obtained mean flow fields around airfoils that were more consistent with experimental measurements than those obtained using the baseline turbulent model. However, the computational time of their approach is high owing to its involving an iterative procedure. The second type are data-driven methods. For example, Liu et al. [11] designed a hybrid deep neural network-based reduced-order model (ROM) for the prediction of transonic buffet flow. This ROM comprises a convolutional neural network (CNN) and a convolutional long short-term memory neural network and they used it to map flow fields in previous and future time steps. The results demonstrated that the error increased over time owing to interpolation loss, especially in the shock and boundary layer regions. This problem has been alleviated by extracting structured-grid-based flow field data using graph neural networks (GNNs) [12–14], as GNN frameworks can appropriately treat arbitrary grids with edges and points. Pfaff et al. [13] developed a GNN-based framework named MeshGraphNets that determines inviscid flow fields over airfoils one to two orders of magnitude faster than CFD simulations. Despite these advantages, GNN technologies are not yet mature and are limited by their theory of graph representations, hardware requirements, and use of distributed solving [15].

Another promising method that can attenuate interpolation loss is mesh transformation [], which converts a grid from Cartesian coordinates to curvilinear coordinates. Wang et al. applied mesh transformation and used a variational autoencoder (VAE) framework and a multilayer perception (MLP) framework for predicting steady flow fields around supercritical airfoils [2]. The VAE effectively captured the representative latent features of flow fields, and the MLP established the relationship between the airfoil shapes and latent features. The mean absolute error and maximum predictive error were less than 0.001 and 0.02, respectively. In addition, Chen and Thuerey [15] exploited the mesh transformation technique to determine flow details near a boundary layer and used a U-net architecture to obtain the steady flow fields of various airfoils under a range of flow conditions. Their results had a mean absolute error of approximately 6e-4. However, despite CNNs having proven effective for predicting complex flows, the accuracy of their predictions of sensitive regions and their generalizability to larger-scale datasets need further improvement.

Vision transformers (ViTs) use a self-attention mechanism to incorporate a large amount of global information and establish long-range relationships, and have recently become a popular framework in the field of computer vision. ViTs pretrained on large-scale datasets are superior to CNNs in terms of the speed–accuracy trade-off. Therefore, in this study, an advanced ViT-based encoder–decoder is devised to establish the complex mapping between the geometric information and physical flow fields. Furthermore, the model is trained by transfer learning on large-scale datasets and by fine-tuning on small datasets to increase their accuracy and generalizability to freestream conditions and scenarios involving novel geometries.

In most existing models, all pixels are trained equally, i.e., no emphasis is placed on local sensitive areas, so it can be challenging for these models to capture significant changes near a shock wave. To overcome this limitation, this study involves the design of four methods to encode a geometric input with various information points that effectively represent feature inputs or emphasize regions near a boundary layer and shock wave. Moreover, a unique loss function is developed based on wavelet transformation or gradient distribution and is used to train the model to improve its predictive accuracy near a shock region.

The remainder of this paper is organized as follows. Section 2 describes the numerical and deep-learning methods. Section 3 presents the comparative results and discussions. Section 4 presents the concluding remarks.

# 2. Method

## 2.1. Overview

The objective of this study is to use a deep neural network to infer RANS solutions for the compressible flow over various supercritical airfoils under a range of freestream conditions. **Figure 1** shows the process flow of the deep-learning-based strategy for steady flow field prediction, which can be considered a classical regression learning task. For freestream condition $\mathcal{I}$ and geometrical information $\mathcal{G}$, the output flow fields $\mathcal{Y}$ can be inferred using a neural-network-based model $f$, i.e., $\mathcal{Y} = f(\mathcal{G}, \mathcal{I})$ via the following four steps.

**1. Data generation.** A dataset of 500 supercritical airfoils with approximately 50 angles of attack (AoAs) at Mach number $Ma = 0.73$ are generated. Each sample includes information on the geometry, mesh, and ground-truth flow fields (section 2.2).

**2. Data preprocessing and postprocessing.** As shown in **Fig. 1**, the input is the geometric information of airfoils and freestream conditions, i.e., the *Ma*, the AoA, and the Reynolds number (*Re*). Mesh transformation is performed to make use of structured networks. Specifically, local univalent transformation is performed to convert the original physical space coordinates into curvilinear coordinates with curved coordinate lines [18], with the mesh in the curvilinear coordinates being uniform and structured. Similarly, the physical flow quantities in the curvilinear coordinate space are re-mapped to the physical space coordinates via an inverse transformation. After mesh transformation, various methods are used to encode the geometric information to select the effective feature inputs for the neural network. The details of the mesh transformation and geometric information encoding are presented in section 2.2.

**3. AI-based predictive model.** A modified ViT-based encoder–decoder architecture is developed to predict the flow fields. The raw inputs of the network are the geometric information and freestream condition, and the outputs are the physical quantities of the structured flow (u-velocity, v-velocity, and pressure). Various loss functions based on wavelet transformation and gradient distribution are introduced to increase the predictive accuracy, especially in the region near the shock wave. The trained model is expected to rapidly predict the velocity and pressure distributions in each cell.

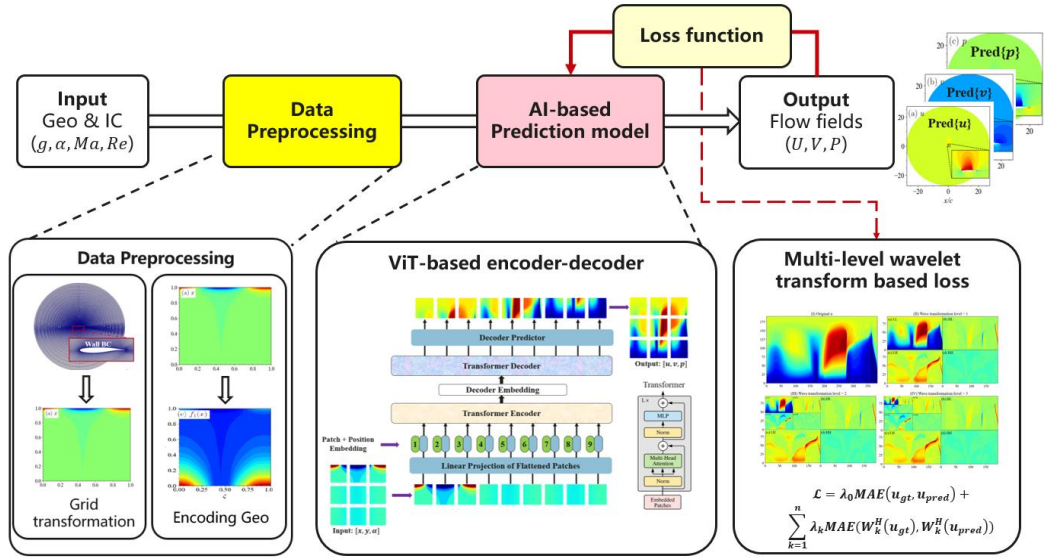

**Fig. 1** Process flow of deep-learning-based approach for flow field prediction

## 2.2. Data generation

**Geometry information:** The research objects are supercritical airfoils applied in engineering design. Two constraints are implied in the data generation process to ensure the diversity and availability of the geometry: the leading-edge radius must be at least 0.007, and the maximum thickness of the airfoil must be within a certain range.

**Mesh information:** As shown in **Fig. 2**, the computational domain is a structured O-grid with a radius of ~30c, where c = 1.0 is the chord length of the airfoil. The grid size normal to the airfoil surface depends on the boundary layer thickness and is rather small within the boundary layer. The grid independence verification results show that the grid has 385 points on the airfoil circumference and 193 points in the normal direction.

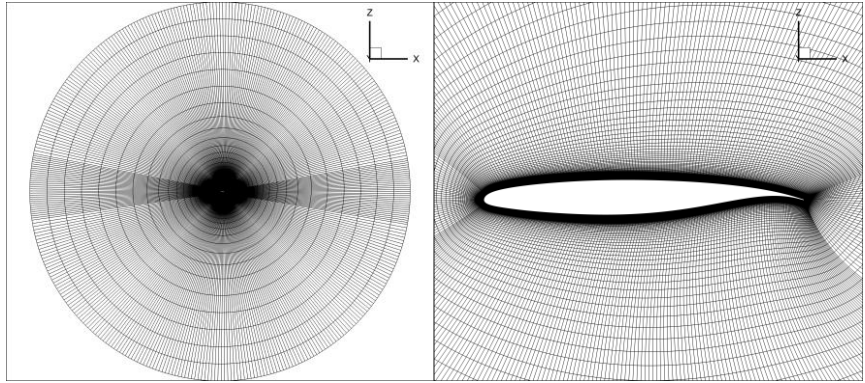

**Fig. 2** Outline and magnified view of the structured mesh adjacent to the airfoil surface.

**Flow information**: RANS simulations use the shear stress transport turbulence model, and solutions are determined using the open-source code CFL3D. The monotonic upstream-centered scheme for conservation laws is used to determine the state-variable interpolations at the cell interfaces, the Roe scheme is used for spatial discretization, and the lower-upper symmetric Gauss–Seidel method is used for time advancement. **Figure 3** shows the pressure distributions on a RAE2822 airfoil obtained from CFD simulations and from experimental tests. As can be seen, the CFD simulations yield accurate results, with the flow field contour indicating that the grid settings can capture the main flow structures.

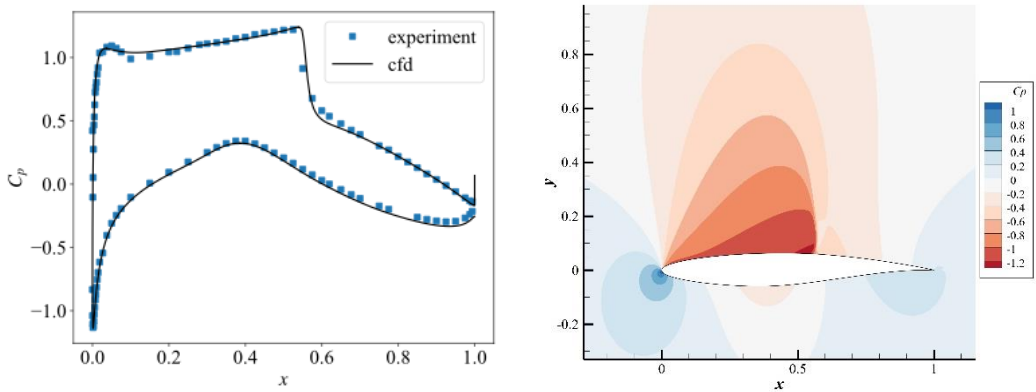

**Fig. 3** Pressure distributions of the RAE2822 airfoil obtained through CFD simulations and experiments.

*2.3. Encoding of geometric information*

The grid size varies across regions because mesh refinement is performed in the boundary layer around the airfoil to retain the flow details. Such data structures cannot be processed by CNN-based or transformer-based neural networks that require uniform and structured inputs. Thus, to provide inputs suitable for these networks and to retain flow information, univalent transformation is performed to convert the geometrical and flow information from Cartesian coordinates $(x, y)$ to curvilinear coordinates $(\xi, \eta)$. Uniform and structured data can be obtained in curvilinear coordinates $(\xi, \eta)$ from the original physical coordinates $(x, y)$ through the following transformation [19,20]:

$$\begin{bmatrix} x \\ y \end{bmatrix}_{i,j} = \begin{bmatrix} x_0 \\ y_0 \end{bmatrix}_{i,0} + \int \begin{bmatrix} dx \\ dy \end{bmatrix} \tag{1}$$

$$\begin{bmatrix} dx \\ dy \end{bmatrix} = \begin{bmatrix} x_\xi & x_\eta \\ y_\xi & y_\eta \end{bmatrix} \begin{bmatrix} d\xi \\ d\eta \end{bmatrix} \tag{2}$$

$$\mathbf{T} = \begin{bmatrix} \xi_x & \xi_y \\ \eta_x & \eta_y \end{bmatrix} = \begin{bmatrix} x_\xi & x_\eta \\ y_\xi & y_\eta \end{bmatrix}^{-1} \tag{3}$$

where

$$\xi = \frac{(i-1)}{(i_{max}-1)}, \quad \eta = \frac{(j-1)}{(j_{max}-1)}; \tag{4}$$

$i, j$ are the indices of the mesh in different directions; $i_{max}$, $j_{max}$ are the number of nodes in these directions, so $i, j$ can be regarded as directions along the body-fitted curve and normal to the body-fitted curve, respectively; $\begin{bmatrix} x_0 \\ y_0 \end{bmatrix}_{i,0}$ is the airfoil profile; and $\mathbf{T}$ is the transformation matrix between these two coordinates. The curvilinear coordinate $(\xi, \eta)$ can be rescaled to a uniform real-number domain $[0,1] \times [0,1]$ using Eq. (4).

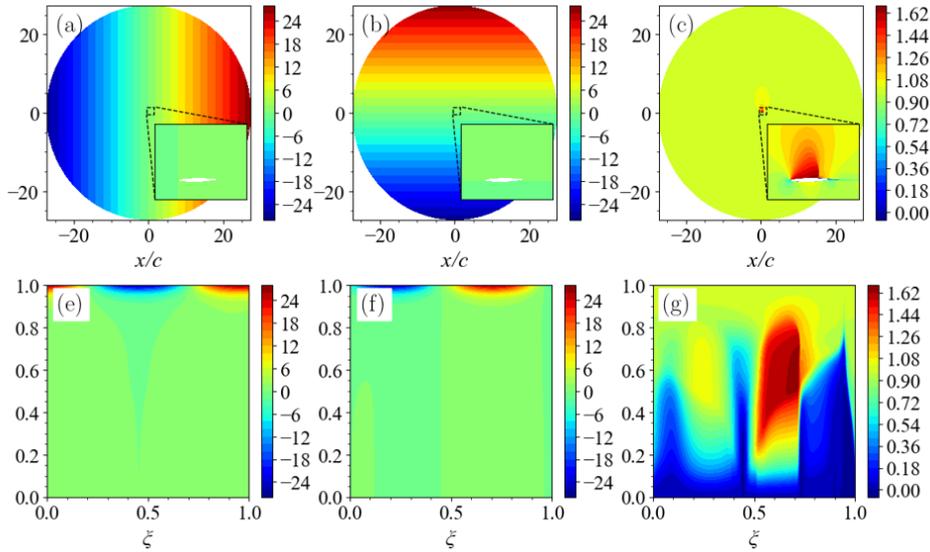

**Fig. 4** Contour plots of various physical quantities before mesh transformation ((a) $x$ map, (b) $y$ map, and (c) $u$ map) and after mesh transformation ((d) $x$ map, (e) $y$ map, and (f) $u$ map).

The contour plots of various physical quantities before and after mesh transformation are shown in **Fig. 4**(a)–(b) and **Fig. 4**(d)–(e), respectively. The detailed flow information near the boundary layer is well preserved in the uniform mesh. However, mesh deformation leads to a distortion between the Cartesian and curvilinear coordinates. To solve this problem, the deformation information, such as the four elements in transformation matrix **T**, are encoded into the input.

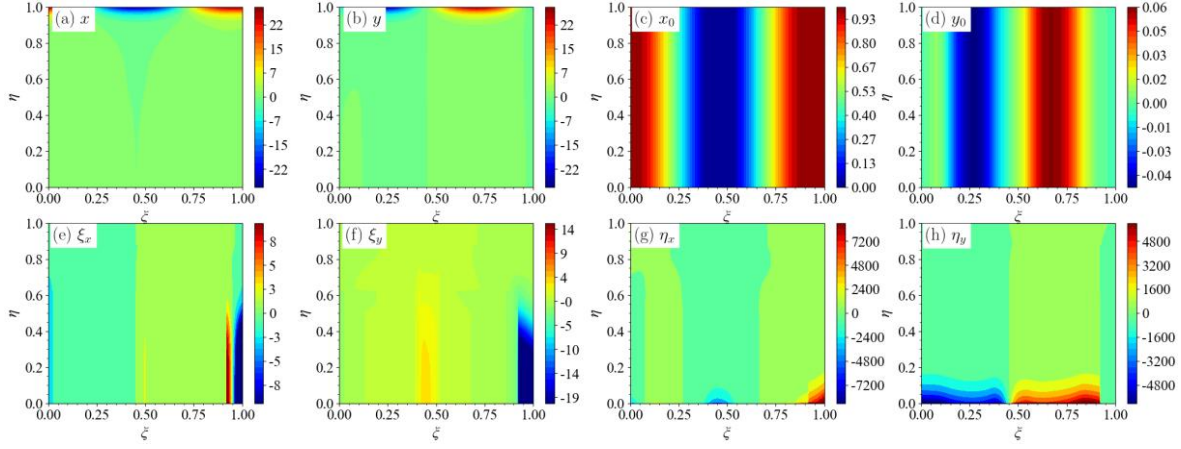

**Fig. 5** Contour plots of various physical quantities after mesh transformation: (a) $x$ map, (b) $y$ map, (c) $x_0$ map, (d) $y_0$ map, (e) $\xi_x$ map, (f) $\xi_y$ map, (g) $\eta_x$ map, and (h) $\eta_y$ map.

With reference to [16], three encoding methods are implemented, as listed in Table 1: Method A, which directly encodes $(x, y)$ to retain the most primitive mesh information; Method B, which encodes the transformation matrix and airfoil profiles $(x_{i,0}, y_{i,0}, \xi_x, \xi_y, \eta_x, \eta_y)$; and Method C, which encodes the original physical coordinates and airfoil profiles $(x_{i,0}, y_{i,0}, x, y)$ to focus on the effect of airfoil profiles. The contour plots are shown in **Fig. 5**. To increase the proportion of flow fields around the airfoils and decrease the proportion of far-field flow, the following filter is designed for method D:

$$M = e^{(-\sigma)}, f_M(x) = M \cdot x, f_M(y) = M \cdot y \tag{5}$$

where $M$ is the function of the filter, and $\sigma$ is the distance to the airfoil surface. As shown in **Fig. 6**, the proportion of near-field flow obtained with this filter is larger than that obtained without this filter.

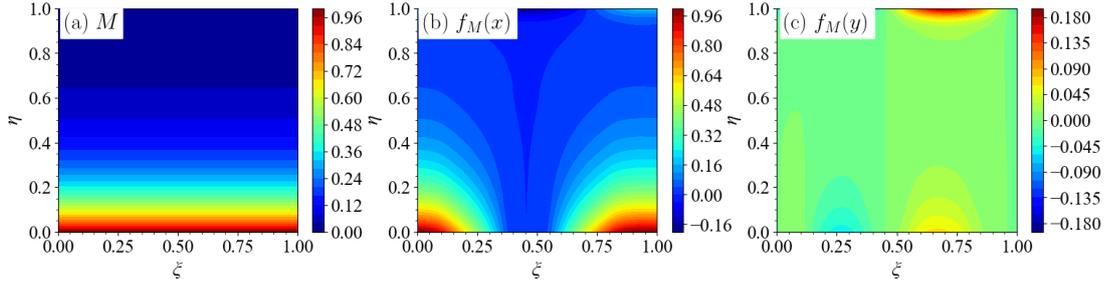

**Fig. 6** Contour plots of (a) filter $M$, (b) an $x$ map after filtering $f_M(x)$, and (c) a $y$ map after filtering $f_M(y)$.

Table 1 Four methods used for encoding the input of the neural network-based model

| Method | Geometric information $\mathcal{G}$ | Freestream condition $\mathcal{J}$ | No. of channels, $C$ |
|---|---|---|---|
| A | $x, y$ | AoA | 3 |
| B | $x_{i,0}, y_{i,0}, \xi_x, \xi_y, \eta_x, \eta_y$ | AoA | 7 |
| C | $x_{i,0}, y_{i,0}, x, y$ | AoA | 5 |
| D | $M \cdot x, M \cdot y$ | AoA | 3 |

## 2.4. Neural network architecture

The deep neural network model used to establish the mapping is based on a modified ViT [17] encoder–decoder, the architecture of which is shown in **Figure 7**. The ViT has a transformer [22] architecture that was developed by Dosovitskiy et al. [17] and has afforded excellent results in natural language processing and computer vision.

In this architecture, the input feature $\mathcal{X} \in \mathbb{R}^{H \times W \times C_{in}}$ is first reshaped into a sequence of two-dimensional (2D) flattened patches $\mathcal{X}_p \in \mathbb{R}^{N_p \times (P^2 \cdot C_{in})}$ through the $Patchify(\mathcal{X}^{H \times W \times C_{in}})$ operation, where $(H, W)$ is the mesh resolution in curvilinear coordinates, $C_{in}$ is the number of input channels, $(P, P)$ is the resolution of each patch, and the number of patches $N_p = HW/P^2$. A trainable linear projection is used to reconstruct the flattened patches to obtain a constant latent vector size $D$, and position embeddings are linearly added to the sequence of patches to retain the positional information. Subsequently, an encoder extracts the effective latent features, and the decoder converts these features into flow field quantities. Both the encoder and decoder contain a transformer mechanism [21] consisting of an attention mechanism and a feedforward neural network. Alternating layers of multihead self-attention [21] and multilayer perception [21] blocks are applied in each transformer encoder, and layernorm and residual connections are applied before and after every block, respectively. The output of the encoder is fed to the decoder, which has the same architecture as the encoder, and the output of the decoder is fed to an MLP block. Subsequently, the "$Unpatchify$" operation (i.e., the inverse of the "$Patchify$" operation) is performed to yield the final output $\mathcal{Y}$ with a size of $H \times W \times C_{out}$.

Because the original ViT is not effective for regression tasks, the following two modifications are introduced to predict flow fields: First, the classification token in the patch and position embeddings is eliminated. Second, a decoder with architecture similar to that of the encoder is used instead of the original MLP. The open-source code MindSpore is applied to design and implement the ViT-based encoder–decoder. The backbone model can be readily replaced by other neural networks, such as U-Net [22]. *The performance evaluation of various models (**Appendix A**) indicates that the ViT-based encoder–decoder is the best baseline model.*

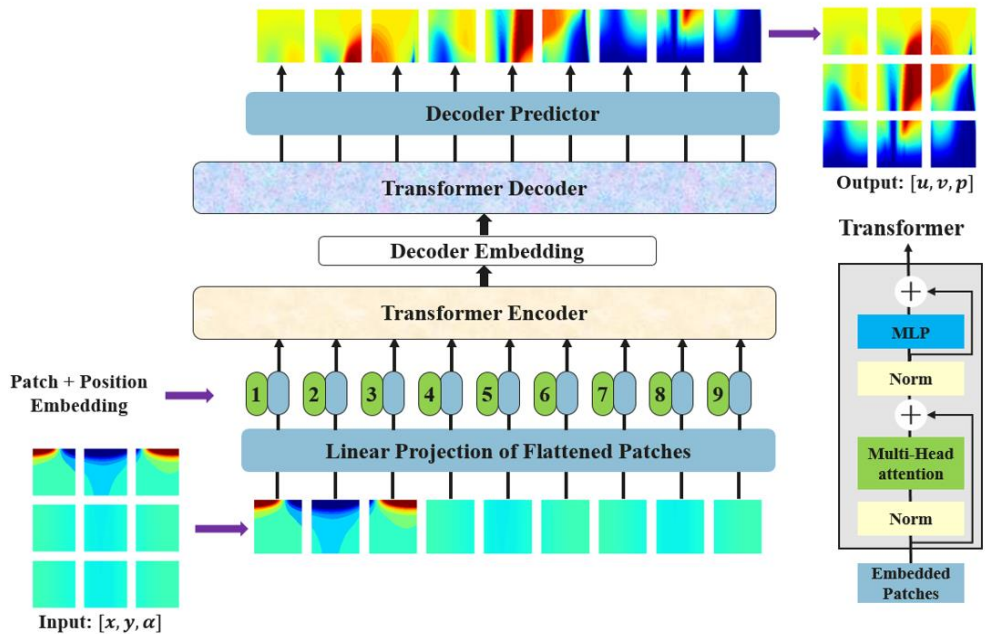

**Fig. 7** Schematic of the architecture of the modified ViT-based encoder–decoder

## 2.5. Loss functions

The mean absolute error (*MAE*) has been typically used to train neural networks for predicting the flow fields over airfoils [2,15]. Therefore, the $l_1$ loss function described in Eq. (6) is used as the baseline loss function in this study, where $N_s$ denotes the number of samples.

$$l_1 = \frac{1}{N_s} \sum_{\{i=1\}}^{N_s} MAE(Y_i^{gt}, Y_i^{pred}) = \frac{1}{N_s} \sum_{\{i=1\}}^{N_s} |Y_i^{gt} - Y_i^{pred}| \quad (6)$$

The above loss function yields satisfactory results over the entire flow field because the same attention is applied to all of the grid nodes in the flow field. However, the physical quantities in different regions of the flow field vary to different extents and thus must be assigned different importance values. For instance, the velocity and pressure vary more near the boundary layer than in the far-field area, which means that that the velocity and pressure are more important in the former area than in the latter area. The flow fields around supercritical airfoils typically involve shock wave or shock wave–boundary layer interactions that exhibit dramatic changes. Few researchers have attempted to increase the predictive accuracy in these local regions. Therefore, in this study, dramatic changes are regarded as high-frequency signals of flow fields, and two loss functions are incorporated into the $l_1$ loss function, i.e., the loss functions of the multilevel wavelet transformation ($l_{wt}$) and gradient distribution ($l_{grad}$).

### 2.5.1 Multilevel wavelet transformation

To define $l_{wt}$, the 2D discrete wavelet transform (DWT) [23] is applied to learn the high-frequency components, as shown in Eq. (7)"

$$l_{wt,k} = \frac{1}{N_s} \sum_{\{i=1\}}^{N_s} |\psi^H(\psi_{k-1}^L(Y_i^{gt})) - \psi^H(\psi_{k-1}^L(Y_i^{pred}))| \quad (7)$$

where $\psi$ is the DWT. $l_{wt,k}$ represents the loss of wavelet transformation at level $k$, where $k$ represents the number of DWT operations. After DWT, four sub-bands can be obtained: the low-frequency component ($LL$) and the high-frequency components in the horizontal direction ($HL$), vertical direction ($LH$), and diagonal direction ($HH$). Thus, $\psi^H$ denotes the high-frequency components after DWT, and $\psi_{k-1}^L$ denotes the low-frequency component after $k-1$ DWT runs. $|.|$ denotes the norm of the discrepancy; both $l_{1,norm}$ and $l_{2,norm}$ are considered in this study. Four convolutional filters are used to realize the 2D DWT [24]. Taking the Haar wavelet [25] as an example, the four convolutional filters corresponding to the four sub-bands are

$$f_{LL} = \begin{bmatrix} 1 & 1 \\ 1 & 1 \end{bmatrix}, f_{LH} = \begin{bmatrix} -1 & -1 \\ 1 & 1 \end{bmatrix}, f_{HL} = \begin{bmatrix} -1 & 1 \\ -1 & 1 \end{bmatrix}, f_{HH} = \begin{bmatrix} 1 & -1 \\ -1 & 1 \end{bmatrix}, \tag{8}$$

The contour plots of the first three levels of the DWT for the streamwise velocity $u$ are shown in **Figs. 8**(II), (III), and (IV). The frequency of the shock wave is reflected in the $HL$ component, and the details of the boundary layers are reflected in the $LH$ component. Compared with the $HL$ and $LH$ components, the values of the $HH$ component are smaller, and the physical representations are unclear. As the DWT level increases, the high-frequency signals in the flow field can be better captured and resolved.

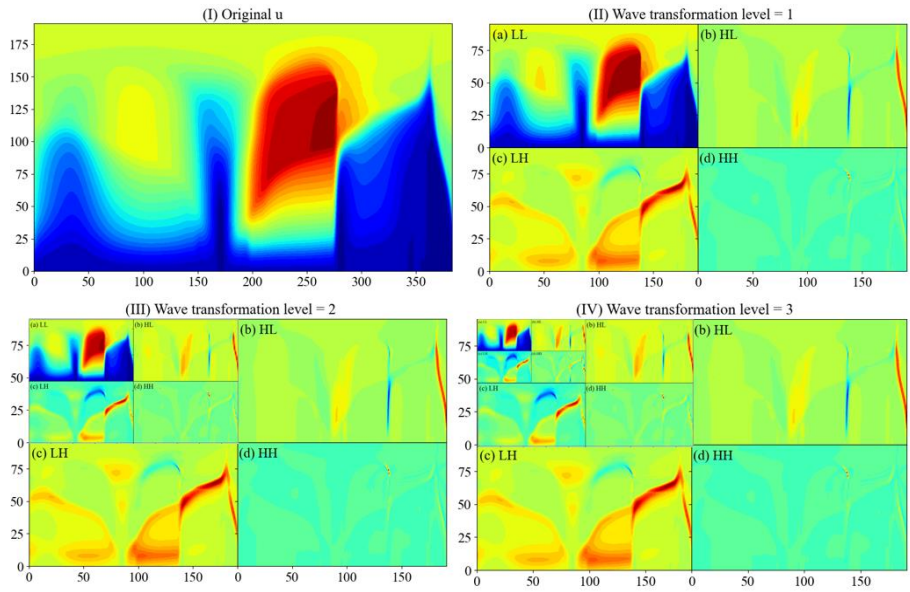

**Fig. 8** Contour plots of different sub-bands obtained using different levels of wavelet transformation

After the DWT implementation, the final form of the total loss function combined with $l_{wt}$ is obtained, as follows:

$$l_{total,wt} = \lambda_0 l_1 + \sum_{k=1}^{n_k} \lambda_k l_{wt,k} \tag{9}$$

where $n_k$ is the number of coupled levels, and $\lambda_0$ and $\lambda_k$ are the weighting coefficients of the $l_1$ loss and wavelet transformation loss at level $k$, respectively. These coefficients are trainable and can be learned through multitask loss optimization [26], as shown in Eq. (10), where $loss_i$ represents the $i_{th}$ loss term, and $\lambda_i$ denotes the corresponding weighting coefficients. The process flow of the ViT-based encoder–decoder trained with $l_{total,wt}$ is shown in **Fig. 9**.

$$l_{total,wt} = \sum_{i=0}^{n} (\frac{loss_i}{2\lambda_i^2} + \log(1 + \lambda_i^2)) \quad (10)$$

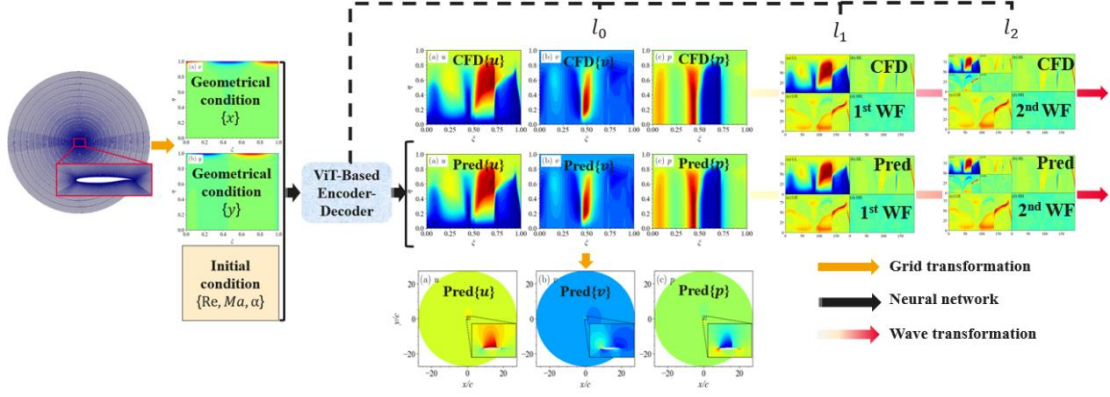

**Fig. 9** Process flow of the ViT-based encoder–decoder trained with $l_{total,wt}$.

### 2.5.2 Gradient distribution

The term $l_{grad}$ is expressed as follows:

$$l_{grad} = \frac{1}{N_s} \sum_{\{i=1\}}^{N_s} |g(y_i^{gt}) - g(y_i^{pred})| \quad (11)$$

where $g$ denotes the gradient in $(\xi, \eta)$ space. The gradient in the $\xi$ and $\eta$ directions can be obtained through the method of central difference. For example, the expressions for the streamwise velocity are as follows:

$$u_\xi(i,j) = \frac{u(i+1,j) - u(i-1,j)}{2}, u_\eta(i,j) = \frac{u(i,j+1) - u(i,j-1)}{2} \quad (12)$$

where $u_\xi$ and $u_\eta$ are the velocity gradients in the $\xi$ and $\eta$ directions, respectively. The contour plots of $u_\xi$ and $u_\eta$ are presented in **Fig. 10**. As can be seen, the spatial distributions of $u_\xi$ and $u_\eta$ are similar to those of the *HL* and *LH* components of DWT, which indicates that the gradient loss can extract high-frequency information from the flow fields. The total loss function is expressed as Eq. (13). The weighting coefficients can be trained in the same manner as in Eq. (10), and the process flow of the ViT-based encoder–decoder trained with $l_{total,grad}$ is illustrated in **Fig. 11**.

$$l_{total,wt} = \lambda_0 l_1 + \lambda_g l_{grad} \tag{13}$$

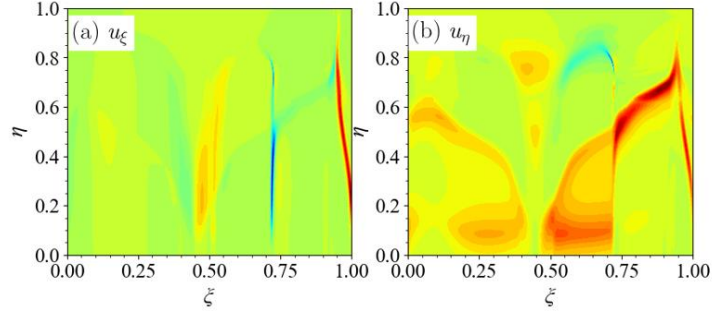

**Fig. 10** Contour plots of the velocity gradient in the $\xi$ and $\eta$ directions.

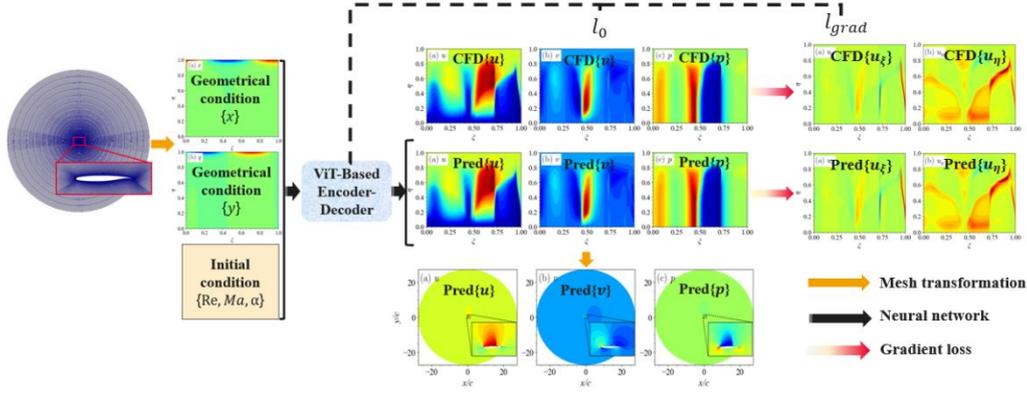

**Fig. 11** Process flow of the ViT-based encoder–decoder trained with $l_{total,grad}$.

## 2.6 Evaluation metrics

The following three error indicators are used to evaluate the model performance.

(1) $l_{1\_avg}$: the difference between the output and original data of each flow field for each sample, defined as

$$l_{1\_avg} = \frac{1}{N_{test}} \sum_{\{i=1\}}^{N_{test}} MAE(y_i^{gt}, y_i^{pred}) \tag{14}$$

(2) $l_{1\_max\_avg}$: the average $l_{1\_max}$ error, defined as

$$l_{1\_max\_avg} = \frac{1}{N_{test}} \sum_{\{i=1\}}^{N_{test}} l_{1\_max,i} = \frac{1}{N_{test}} \sum_{\{i=1\}}^{N_{test}} MAXE(y_i^{gt}, y_i^{pred}) \tag{15}$$

(3) $l_{1\_max\_max}$: the maximum $l_{1\_max}$ error of the testing samples, defined as

$$l_{1\_max\_max} = MAXE\{MAXE(y_i^{gt}, y_i^{pred}) \,|\, i = 1, \dots, N_{test}\} \tag{16}$$

where $MAE$ and $MAXE$ are the mean absolute error and maximum absolute error, respectively.

## 3. Results and discussion

The ability of the devised method to estimate the velocity and pressure fields around various airfoils under a range of flow conditions is evaluated. First, the performances of the different encoding methods are compared in terms of their ability to obtain the effective input features. Subsequently, different loss functions are applied to enhance the predictive accuracy, especially regarding the shock wave.

### *3.1. Comparison of encoding methods*

The four encoding methods listed in Table 1 are used to obtain the effective geometrical information. The testing $l_{1\_avg}$ error and $l_{1\_max\_avg}$ error curves of the ViT-based encoder–decoder obtained from these encoding methods with respect to the epochs are shown in **Fig. 12**. The convergence values of $l_{1\_avg}$, $l_{1\_max\_avg}$, and $l_{1\_max\_max}$ are summarized in **Table 2**. The testing errors decrease rapidly in the first 400 epochs and then decrease steadily from epoch 400 to 800. The curve of the initial descent for encoding method A is small because the initial learning rate is sufficiently small to prevent training divergence. After training for ~1,000 epochs, the testing errors $l_{1\_avg}$ and $l_{1\_max\_avg}$ converge at magnitudes of ~$10^{-4}$ and ~$10^{-2}$, respectively.

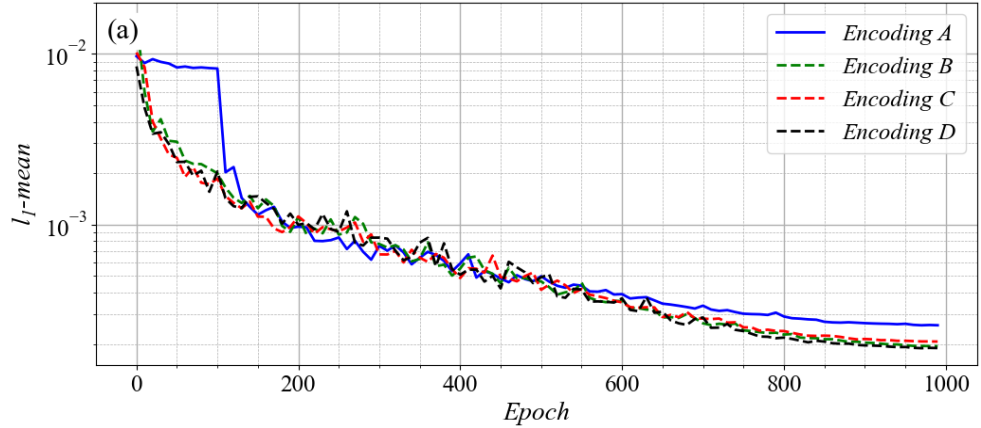

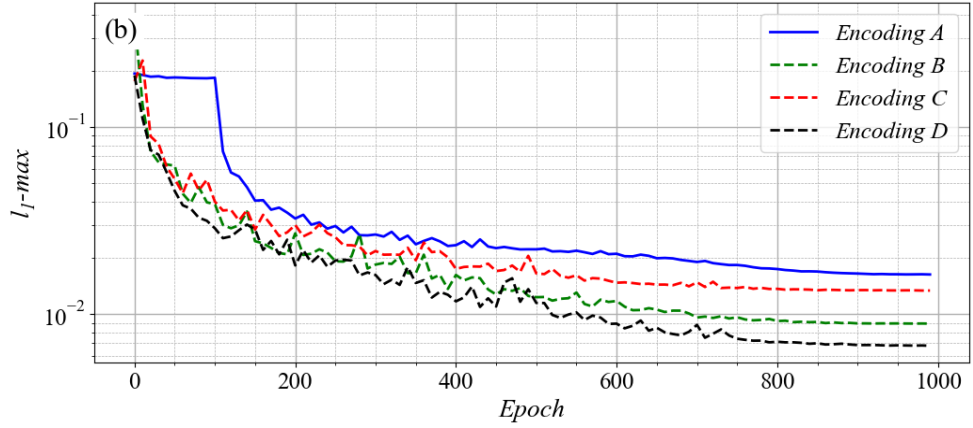

**Fig. 12** Testing error (a) $l_{1\_avg}$ and (b) $l_{1\_max\_avg}$ curves of various encoding methods

**Table 2** Comparison of $l_{1\_avg}$, $l_{1\_max\_avg}$, and $l_{1\_max\_max}$ of ViT-based encoder–decoder using various encoding methods

| Method | $l_{1\_avg}$ | $l_{1\_max\_avg}$ | $l_{1\_max\_max}$ |
| --- | --- | --- | --- |
| A | 0.000258 | 0.0163 | 0.0981 |
| B | 0.000210 | 0.01750 | 0.0956 |
| C | 0.000206 | 0.01341 | 0.08231 |
| D | 0.000187 | 0.0071 | 0.0687 |

The distribution statistics of the $l_{1\_avg}$ and $l_{1\_max}$ errors for the flow fields predicted using various encoding methods are depicted in **Fig. 13**. **Fig. 13**(a) shows that the $l_{1\_avg}$ errors of most samples less than 4e-4, with method A exhibiting the worst performance (i.e., the maximum error). Method D exhibits the best performance (i.e., the minimum $l_{1\_max}$ error). Two apparent peaks can be observed in **Fig. 13**(b) that represent the occurrence of a shock wave. Overall, method D captures the shock region with the minimum error: its $l_{1\_max\_avg}$ and maximum error $l_{1\_max\_max}$ are 58% and 40% smaller than those of other methods, respectively.

Moreover, compared with method A, methods B and C exhibit a smaller $l_{1\_avg}$ error and slightly larger $l_{1\_max\_avg}$ error. This indicates that consideration of additional geometrical information (such as the Jacobin matrix or airfoil profiles) can enhance the accuracy of the overall flow fields but cannot decrease the maximum error in several local areas. Method D achieves the minimum $l_{1\_max\_avg}$ error owing to the enlargement of the near-field flow, which helps the neural network to learn the regions near the boundary layer and shock wave that have the greatest discrepancies.

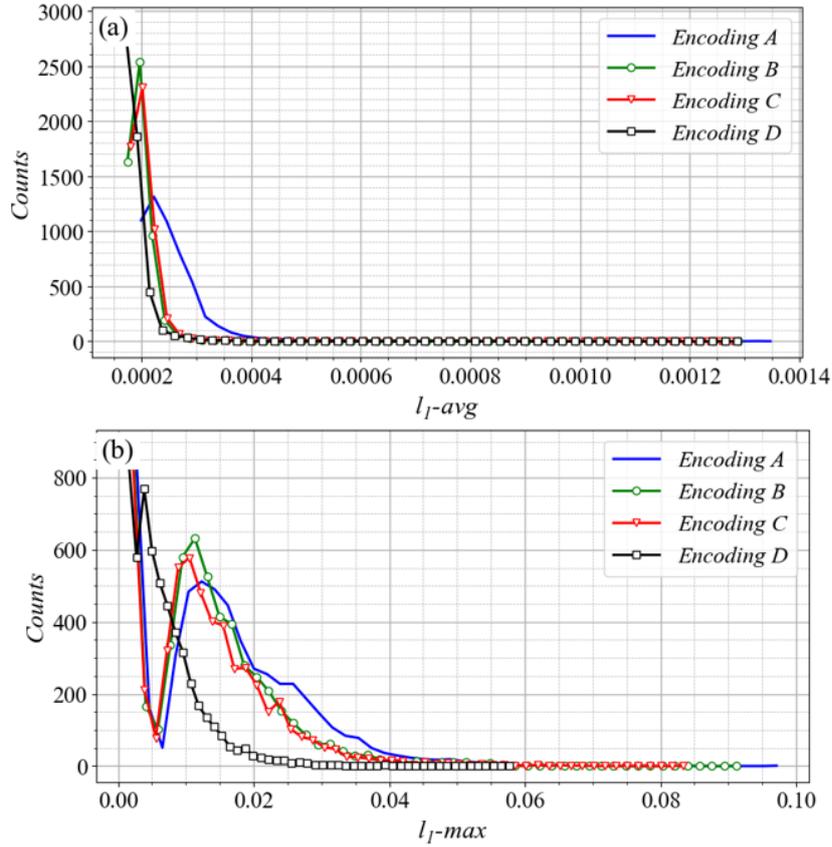

**Fig. 13** Distribution statistics of (a) $l_{1\_avg}$ and (b) $l_{1\_max\_avg}$ for the predicted along-surface flow fields

Four testing samples, three of which contain shock phenomena, are randomly selected to demonstrate the performance of the model. The flow fields predicted using different encoding methods (A, C and D), and their corresponding $l_1$ errors with respect to the ground truth, are presented in **Figs. 14–19**. The flow fields predicted using method B are not presented because they are comparable to those predicted using method C. The flow fields without the shock wave are similar, and nearly all of the regions are similarly predicted. The predicted contour patterns of the flow fields with the shock wave are consistent with the CFD results, although apparent discrepancies can be observed near the shock. This result is reasonable because the neural network can more easily capture low-frequency signals than high-frequency signals. Overall, the flow fields predicted by all of the methods are consistent with the ground truth data.

Compared with the results obtained using methods A and C, the results obtained using method D are in better agreement with the CFD results, and the detailed flow structures near the shock wave are more precisely captured. Therefore, method D obtains the minimum $l_{1\_max}$ error, which indicates that effective feature selection enhances predictive accuracy in high-frequency areas. The number of channels in method D is the same as that in method A, which indicates that no additional computational cost is incurred by method D.

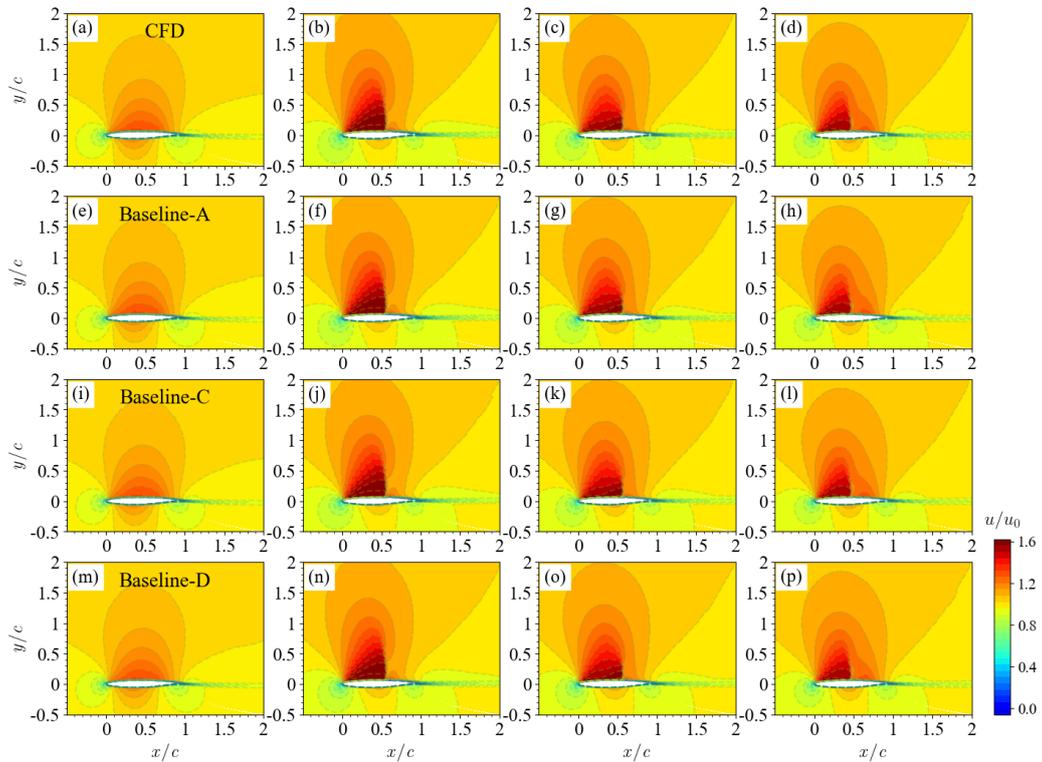

**Fig. 14** Contour plots of non-dimensional streamwise velocity of various airfoils determined using (a)–(d) CFD (GT); (e)–(h) encoding method A; (i)–(l) encoding method C; and (m)–(p) encoding method D.

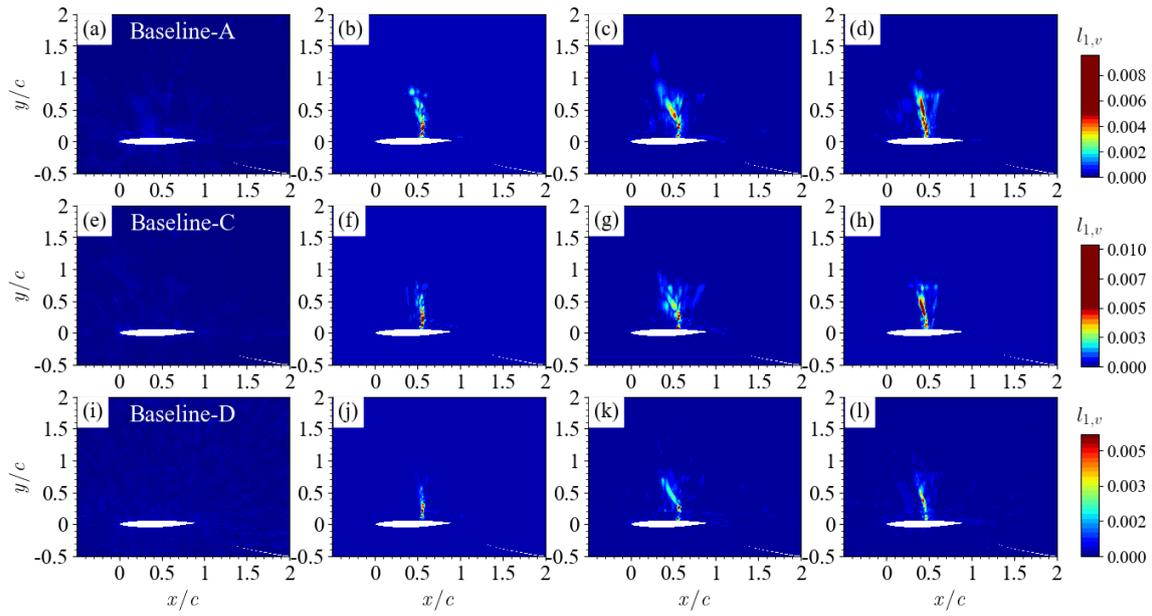

**Fig. 15** Contour plots of $l_1$ error map of streamwise velocity of various airfoils determined using (a)–(d) encoding method A; (e)–(h) encoding method C; and (i)–(l) encoding method D.

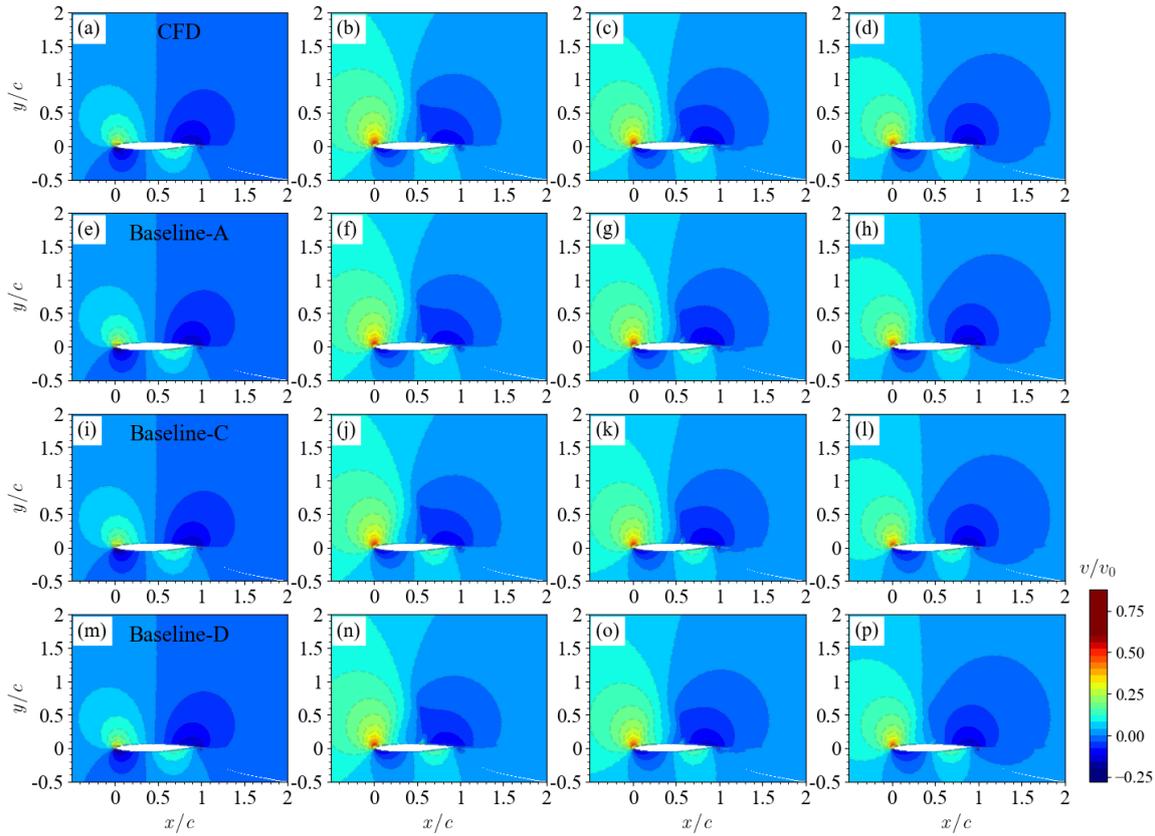

**Fig. 16** Contour plots of non-dimensional normal velocity of various airfoils determined using (a)–(d) CFD (GT); (e)–(h) encoding method A; (i)–(l) encoding method C; and (m)–(p) encoding method D.

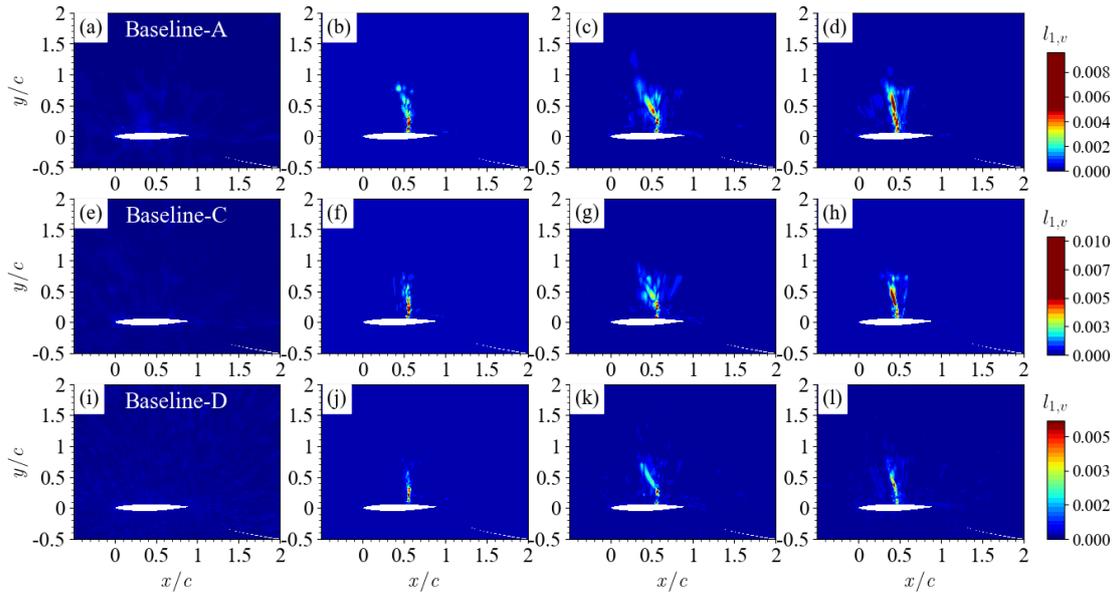

**Fig. 17** Contour plots of $l_1$ error map of normal velocity of various airfoils determined using (a)–(d) encoding method A; (e)–(h) encoding method C; and (i)–(l) encoding method D.

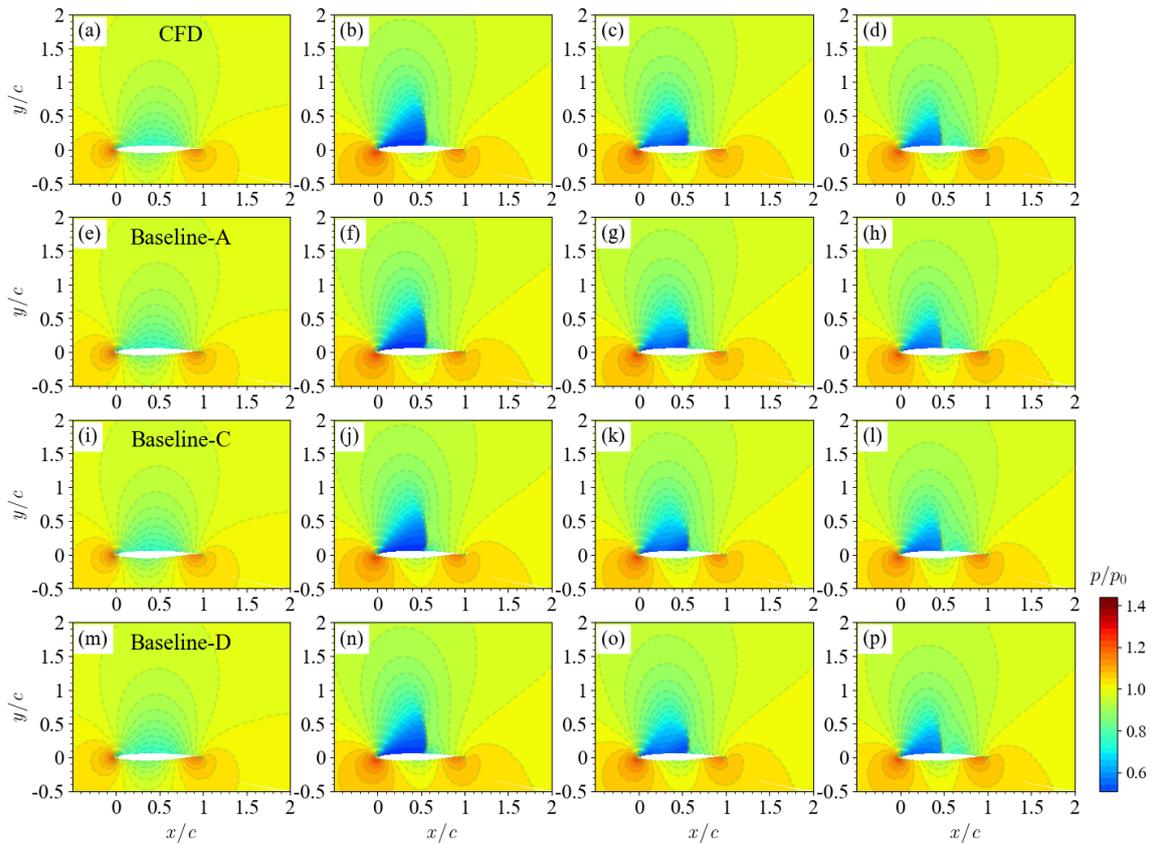

**Fig. 18** Contour plots of non-dimensional pressure of various airfoils determined using (a)–(d) CFD (GT); (e)–(h) encoding method A; (i)–(l) encoding method C; and (m)–(p) encoding method D.

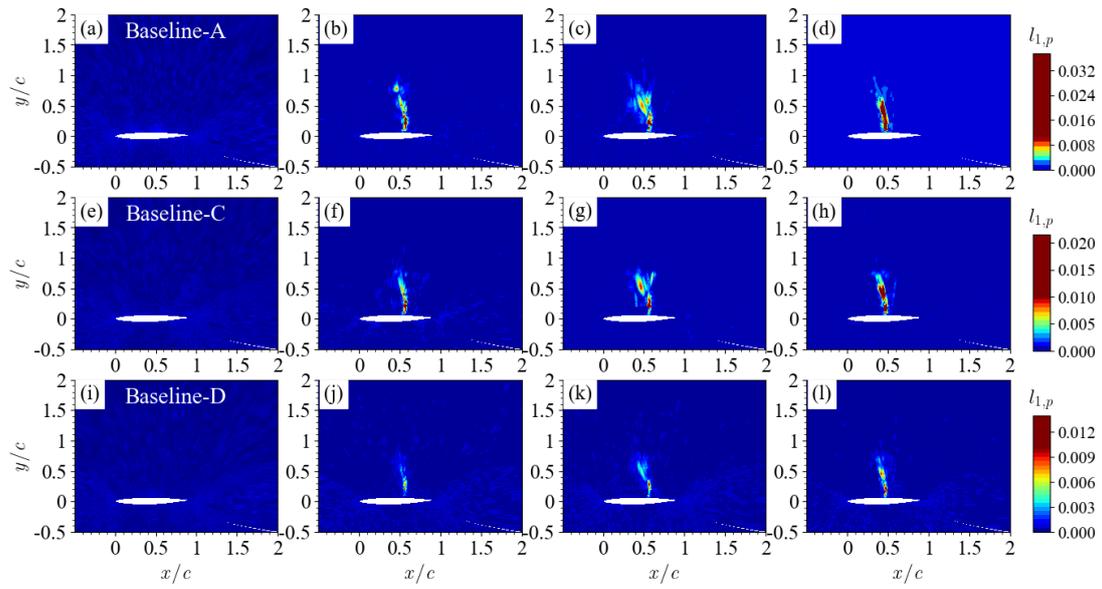

**Fig. 19** Contour plots of $l_1$ error map of pressure of various airfoils determined using (a)–(d) encoding method A; (e)–(h) encoding method C; and (i)–(l) encoding method D.

## 3.2. Comparison of loss functions

Although the strategy with different encoding methods can effectively predict the overall flow fields, the accuracy near the shock wave region must be increased. This section evaluates the two loss functions introduced in section 2.5.

The $l_{1\_avg}$, $l_{1\_max\_avg}$, and $l_{1\_max\_max}$ errors of various loss functions with different encoding methods are presented in **Table 3**. The errors for all four encoding methods decrease with the addition of the gradient loss or the wavelet loss. The $l_{1\_max\_avg}$ error exhibits the highest reduction, as expected, given that the shock wave region can be effectively captured by this maximum error. The modified loss function decreases the $l_{1\_max\_avg}$ error by 50% for methods A, B, and C and by 30% for method D. This indicates that the modified loss function enhances the predictive accuracy in both the overall flow fields and in several local areas, especially in scenarios involving shock wave phenomena.

The degree of reduction in $l_{1\_max\_max}$, which is the maximum error of all of the samples, varies between methods A, B, C, and D, which indicates that the neural networks can assign adequate attention to a single outlier.

Table 3 Comparison of various loss functions

| Method | loss func | $l_{1\_avg}$ | | $l_{1\_max\_avg}$ | | $l_{1\_max\_max}$ | |
|---|---|---|---|---|---|---|---|
| A | $l_1$ | 0.000258 | - | 0.0163 | | 0.0981 | |
| | $l_1 + l_{grad}$ | 0.000186 | 27.9% | 0.0083 | 49.1% | 0.0884 | 9.9% |
| | $l_1 + l_{wt}$ | 0.000194 | 24.8% | 0.0080 | 50.9% | 0.0770 | 21.5% |
| B | $l_1$ | 0.000210 | - | 0.01750 | | 0.0956 | |
| | $l_1 + l_{grad}$ | 0.000169 | 19.5% | 0.00636 | 63.7% | 0.0338 | 64.6% |
| | $l_1 + l_{wt}$ | 0.000180 | 14.3% | 0.00517 | 70.5% | 0.0410 | 57.1% |
| C | $l_1$ | 0.000206 | - | 0.01341 | | 0.08231 | |
| | $l_1 + l_{grad}$ | 0.000167 | 18.9% | 0.00538 | 59.9% | 0.0569 | 30.9% |
| | $l_1 + l_{wt}$ | 0.000187 | 9.2% | 0.00592 | 55.9% | 0.0321 | 61.0% |
| D | $l_1$ | 0.000187 | - | 0.00714 | | 0.0687 | |

|  |  |  |  |  |  |  |
|---|---|---|---|---|---|---|
| $l_1 + l_{grad}$ | 0.000162 | 13.6% | 0.00384 | 46.2% | 0.0475 | 30.8% |
| $l_1 + l_{wt}$ | 0.000171 | 8.8% | 0.00430 | 39.7% | 0.0612 | 10.8% |

To investigate the effect of different loss functions, the predicted flow fields are further investigated by considering method B as the benchmark (as method B gives the greatest discrepancy). In general, increasing the level of wavelet transformation incurs additional computational costs that may outweigh the benefit of loss minimization. Therefore, only first-order (wave loss 1), second-order (wave loss 2), and third-order (wave loss 3) wavelet transformations are applied in this study. The corresponding $l_{1\_avg}$ and $l_{1\_max\_avg}$ errors of various loss functions with respect to the epochs are plotted in **Fig. 20**. The $l_{1\_max\_avg}$ error decreases by an order of magnitude after the addition of the gradient loss or the wavelet loss, and the second-order wavelet transformation loss is the most effective, as it has the lowest $l_{1\_max\_avg}$ error.

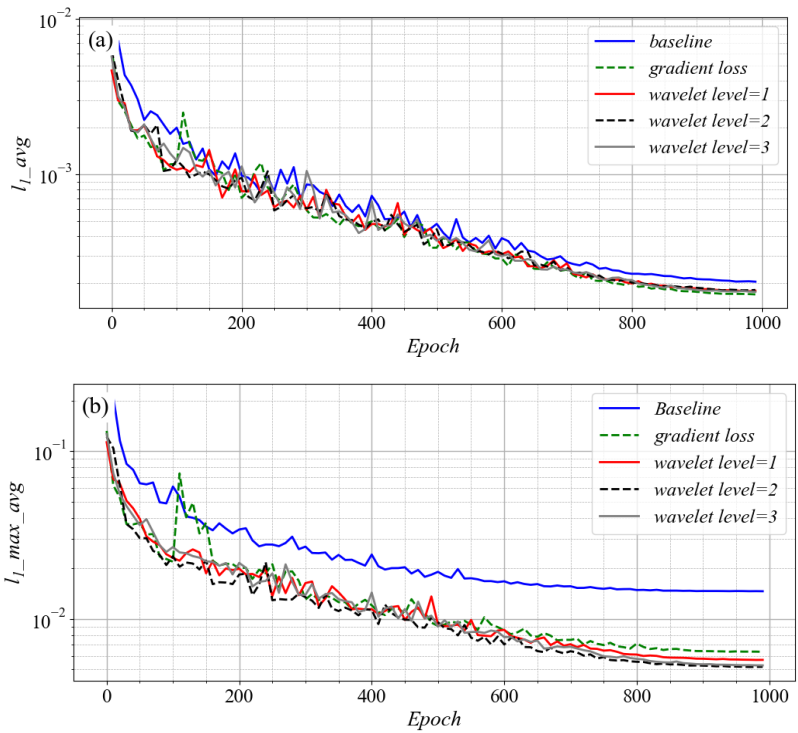

**Fig. 20** (a) $l_{1\_avg}$ error and (b) $l_{1\_max\_avg}$ error of various loss functions with respect to epochs using various loss functions

The distribution statistics of the $l_{1\_max}$ error for the flow fields predicted using different loss functions are presented in **Fig. 21**. The $l_{1\_max}$ errors of most samples in the baseline are distributed in two intervals ([0, 0.0035], [0.0075, 0.03]), depending on the shock phenomena. After the addition of the gradient loss or the wavelet loss, the errors decrease, with most values lying in the interval [0, 0.0065]. In addition, the wavelet loss is slightly more effective than the gradient loss, given its lower $l_{1\_avg}$ error and $l_{1\_max\_avg}$ error.

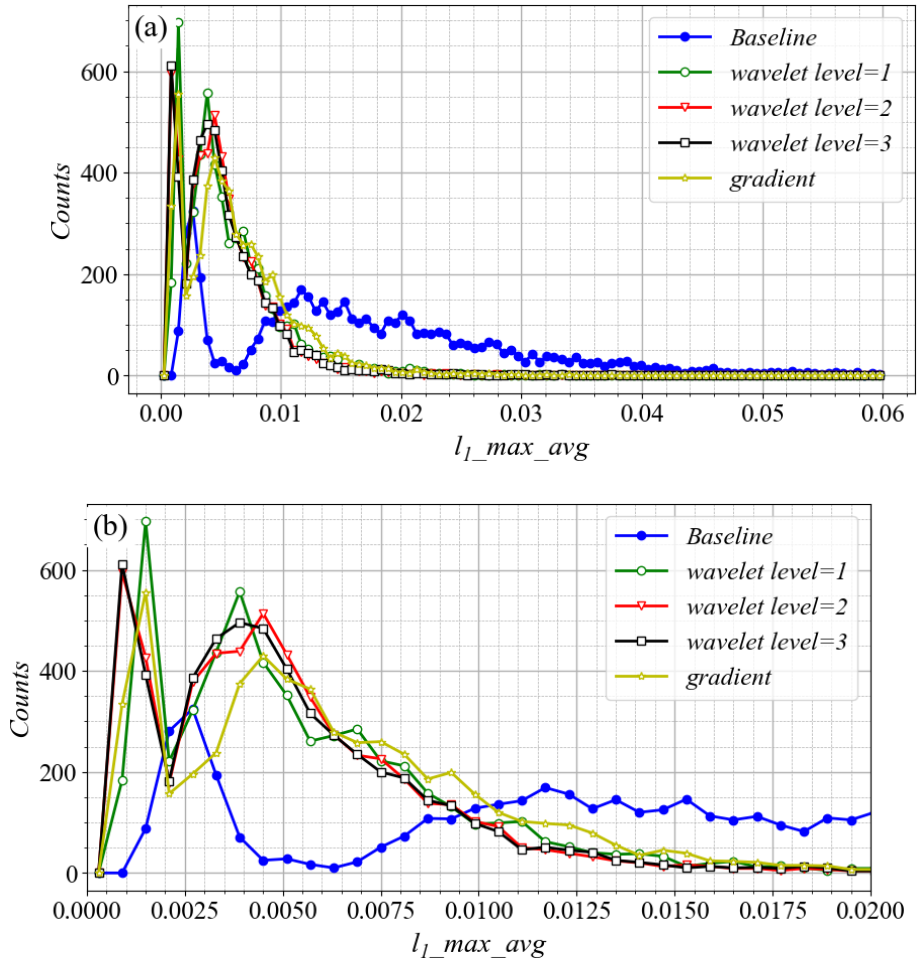

**Fig. 21** (a) Overall and (b) expansion of distribution statistics of the $l_{1\_max\_avg}$ error for the predicted flow fields (density, general bin)

Four testing samples are randomly selected to demonstrate the performance of different loss functions. The flow fields predicted using different encoding methods (A, C, and D), and their corresponding errors with respect to the ground truth, are presented in **Figs. 22–27**, respectively. As can be seen, the details of the velocity and pressure fields are faithfully predicted, even for complicated flow scenarios with shock waves. The predictive accuracy of the flow fields without a shock wave is reasonable. In the flow fields with a shock wave, there remains a discrepancy in comparison with the CFD results in the region near the shock. However, after adding the gradient loss or the wavelet loss, the maximum $l_1$ error decreases from 0.04 to 0.02 or 0.01; this indicates that the shock wave region is more precisely captured when either of these losses are considered than when they are not. Overall, the introduction of the high-frequency component into the flow fields with loss functions decreases the maximum error in several local areas, such as the maximum error in the prediction of the velocity and pressure near the shock wave.

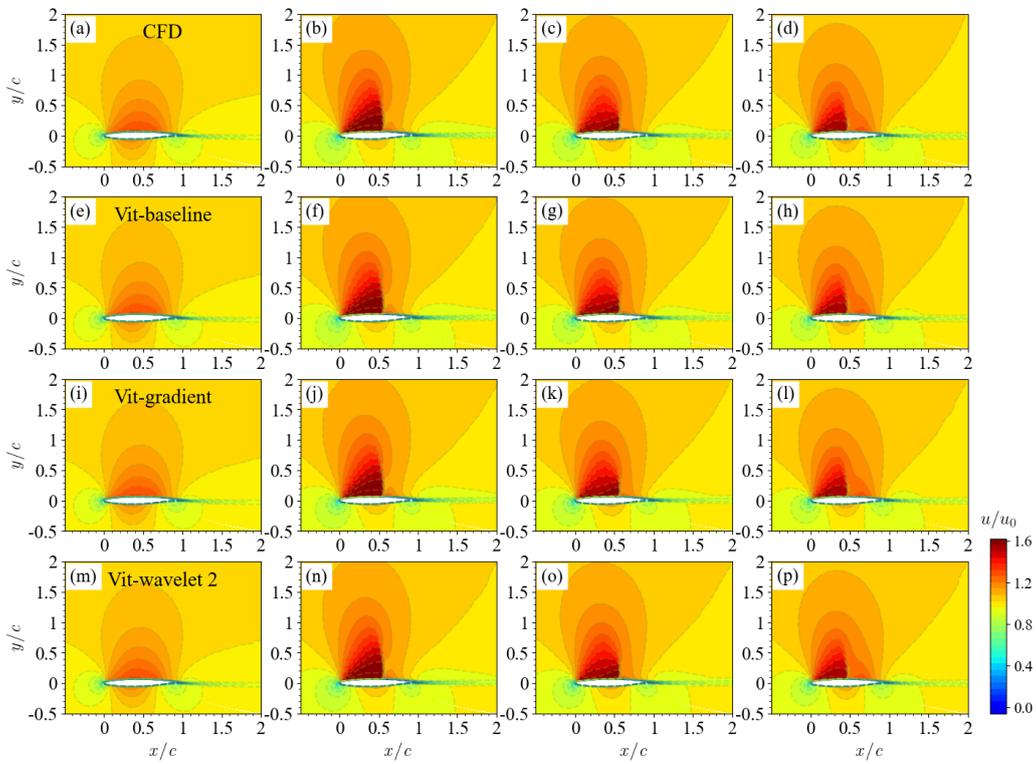

**Fig. 22** Contour plots of non-dimensional streamwise velocity of various airfoils determined using (a)–(d) CFD; (e)–(h) a ViT model trained with the $l_1$ loss; (i)–(l) a ViT model trained with the gradient loss; and (m)–(p) a ViT model trained with the second-level wavelet loss.

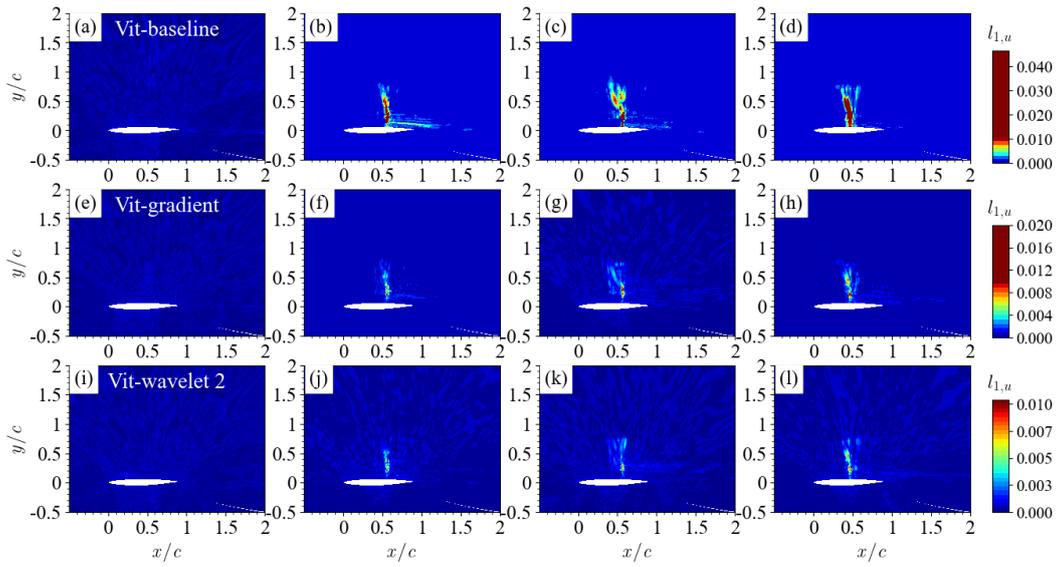

**Fig. 23** Contour plots of $l_1$ error map of streamwise velocity of various airfoils determined using (a)–(d) CFD; (e)–(h) a ViT model trained with the $l_1$ loss; (i)–(l) a ViT model trained with the gradient loss; and (m)–(p) a ViT model trained with the second-level wavelet loss.

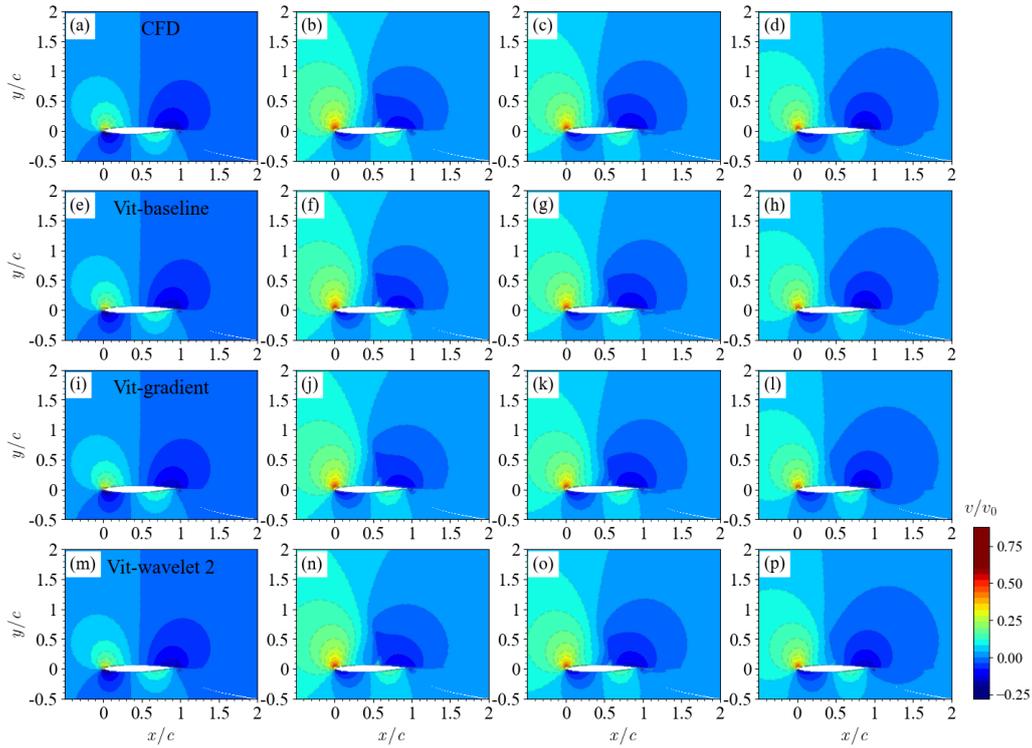

**Fig. 24** Contour plots of non-dimensional normal velocity of various airfoils determined using (a)–(d) CFD; (e)–(h) a ViT model trained with the $l_1$ loss; (i)–(l) a ViT model trained with the gradient loss; and (m)–(p) a ViT model trained with the second-level wavelet loss.

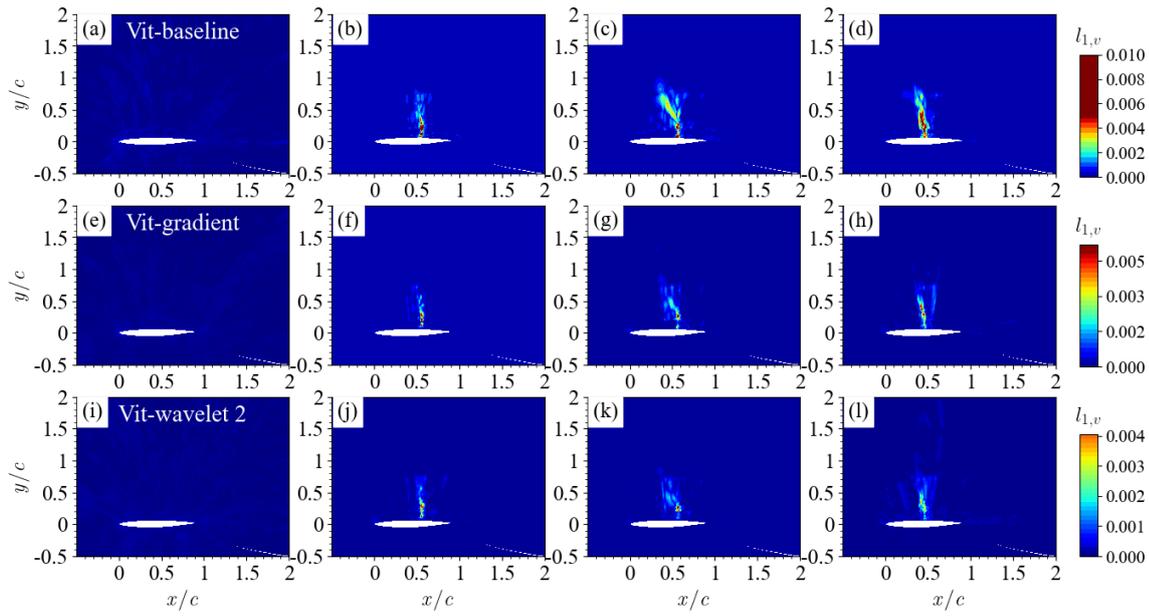

**Fig. 25** Contour plots of $l_1$ error map of normal velocity of various airfoils determined using (a)–(d) CFD; (e)–(h) a ViT model trained with the $l_1$ loss; (i)–(l) a ViT model trained with the gradient loss; and (m)–(p) a ViT model trained with the second-level wavelet loss.

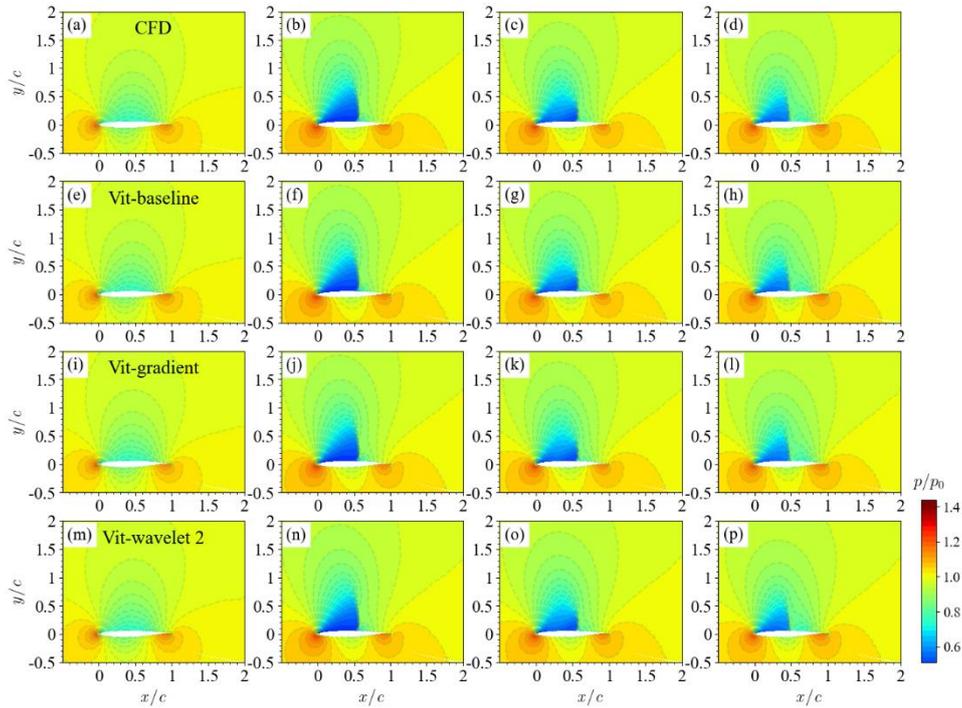

**Fig. 26** Contour plots of non-dimensional pressure of various airfoils determined using (a)–(d) CFD; (e)–(h) a ViT model trained with the $l_1$ loss; (i)–(l) a ViT model trained with the gradient loss; and (m)–(p) a ViT model trained with the second-level wavelet loss.

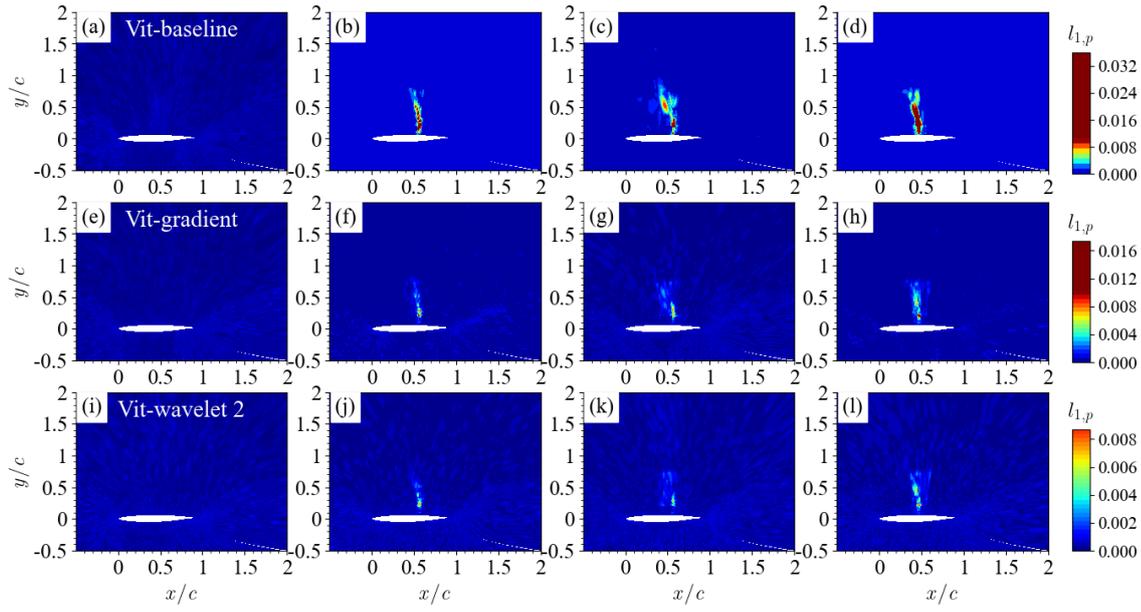

**Fig. 27** Contour plots of $l_1$ error map of the pressure of various airfoils determined using (a)–(d) CFD; (e)–(h) a ViT model trained with the $l_1$ loss; (i)–(l) a ViT model trained with the gradient loss; and (m)–(p) a ViT model trained with the second-level wavelet loss.

To depict the flow details, the predicted shock-wave positions are compared. The shock-wave position is typically determined based on the maximum gradient of pressure in the streamwise direction. As shown in **Fig. 28**, the results obtained from CFD and those obtained using the developed strategy are similar, indicating that the shock wave position is accurately captured by the strategy. The pressure coefficient ($C_p$) distributions along the airfoil surface of randomly selected samples are shown in **Fig. 29**. The $C_p$ values for various flow configurations are consistent with the CFD results, which demonstrates that the strategy developed in this study accurately captures local flows.

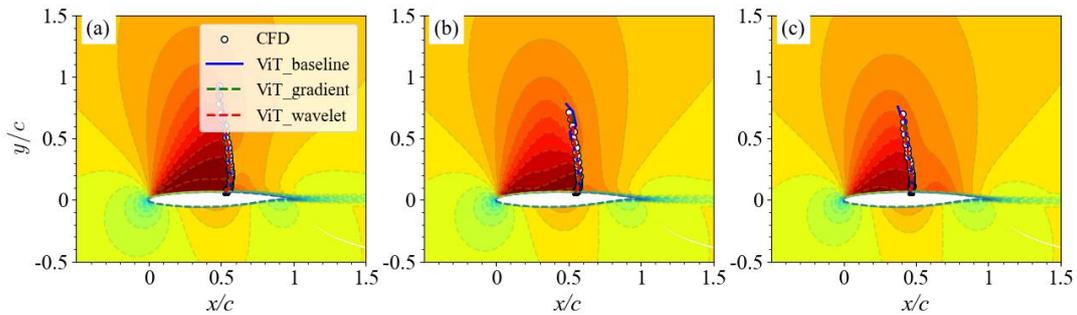

**Fig. 28** Comparison of the shock wave positions for various airfoils determined using CFD (white dot),

ViT_baseline (blue solid line), ViT_gradient (green dashed line), and ViT_wavelet (red dashed line).

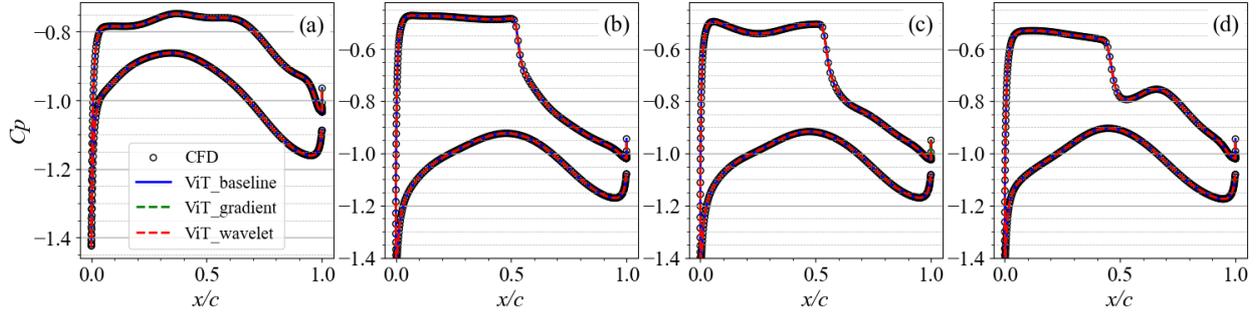

**Fig. 29** Comparison of the surface pressure coefficient ($C_p$) for various airfoils determined using CFD (white dot), ViT_baseline (blue solid line), ViT_baseline (green dashed line), and ViT_wavelet (red dashed line).

The wavelet loss is slightly superior to the gradient loss in these experiments, although there is a similar decrease in the maximum error achieved using both loss functions. This comparable performance is attributable to the fact that both loss functions help the neural network to precisely capture the high-frequency component, which includes the signals of the shock wave. Although the diagonal high-frequency component is introduced into the wavelet transformation, it is incidental compared with the horizontal and vertical components. Although the error near the shock wave cannot be eliminated, the shock-wave predictive accuracy of the developed strategy is higher than that achieved by previously reported strategies. In terms of the time required for the inference, 5,500 flow fields can be obtained within 72 s by a pretrained model running on a single Ascend 910 platform, which is hundreds of times faster than the traditional solver. These results demonstrate that the developed strategy rapidly and accurately predicts the flow over a supercritical airfoil.

*3.3. Transfer learning for engineering applications*

When designing a new airfoil, various inflow conditions (such as AoA or *Ma* values) must be simulated for comprehensive performance evaluation. To improve the generalizability of the model and thereby increase its utility in engineering scenarios, transfer learning is used to pre-train the model on large-scale datasets and fine-tune the model on small datasets. The configurations of the pre-training and fine-tuning datasets are presented in **Table 4**. Four scales of the dataset are set to pre-train the model to determine the minimum number of training samples required for engineering applications, with consideration of the trade-off between precision and time consumption (i.e., compared with pre-training over a smaller dataset, pre-training over a larger dataset may yield more accurate results but have a higher computational time). Few-shot transfer learning is applied to fine-tune the model parameters, as this process requires only 0–5 snapshots for a new airfoil design with various AoAs. The flow fields for the other inflow conditions of the airfoil can be directly inferred using the fine-tuned model.

Table 4 Various configurations of the pre-training and fine-tuning datasets

| Pre-training dataset | Airfoil index | Number of samples | Fine-tuning dataset | Inference dataset |
|---|---|---|---|---|
| Tiny | 0–50 | 2,805 | Airfoil index: 451–500 Snapshots: [0, 1, 3, 5] | $33 \times 50 = 1,650$ |
| Small | 0–100 | 5,534 | | |
| Medium | 0–250 | 13,756 | | |
| Large | 0–450 | 24,744 | | |

The results of transfer learning are presented in **Table 5**. When the model is pretrained using the tiny dataset, at least three snapshots are required to achieve a precision of 4e-4. In contrast, when the model is pre-trained using the small, medium, or large datasets, only one snapshot is required, and an accuracy of 1e-4 can be maintained. Additionally, $l_{1\_avg}$ can be reduced by at least 50% through transfer learning with five snapshots. The model pretrained using the large dataset can predict the flow fields with a certain precision without fine-tuning. The fine-tuning results obtained using datasets of different scales and different numbers of snapshots are shown in **Fig. 30**. The time required for fine-tuning is considerably lower than that for generating samples. Therefore, fine-tuning techniques are particularly valuable for use in engineering applications.

Table 5 Results of the pretrained model fine-tuned using various numbers of snapshots

| Pre-training dataset | Airfoil index | $l_{1\_avg}$ of the inference dataset | | | |
|---|---|---|---|---|---|
| | | 0-shot | 1-shot | 3-shot | 5-shot |
| Tiny | 0–50 | 0.002734 | 0.0012067 | 0.000644 | 0.000448 |
| Small | 0–100 | 0.001831 | 0.0007356 | 0.0004237 | 0.000317 |
| Medium | 0–250 | 0.001077 | 0.000439 | 0.0002798 | 0.000224 |
| Large | 0–450 | 0.000377 | 0.0002388 | 0.00020 | 0.000182 |
| Time consumption (s) | | - | 164 | 176 | 182 |

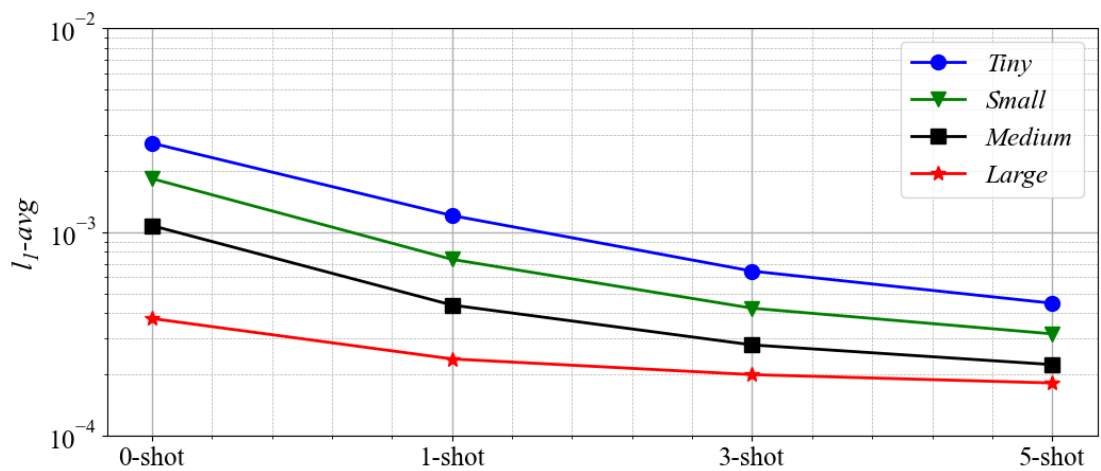

**Fig. 30** Comparison of fine-tuning results obtained using datasets with various scales and numbers of snapshots

## 4. Concluding remarks

This paper devises an advanced ViT-based encoder–decoder network to rapidly and accurately predict steady flow fields over supercritical airfoils under various freestream conditions. To improve the accuracy near regions with drastic changes, the geometric input is encoded with various information points, and the loss functions are modified with a gradient distribution and a multilevel wavelet transformation. Transfer learning is applied to enhance the generalizability of the model trained on small datasets to engineering applications, in which the model is trained on large-scale datasets and fine-tuned on small datasets. Training is performed using a dataset consisting of more than 27,000 flow fields over approximately 500 supercritical airfoils with 50 AoAs, and is simulated using RANS modeling. The following conclusions are derived.

1) Four methods are employed to encode the geometric input: method A, which uses primitive mesh information; method B, which uses the transformation matrix and airfoil coordinates; method C, which uses primitive mesh information and airfoil coordinates; and method D, which uses filtered primitive mesh information. The consideration of additional geometrical information (such as the transformation matrix or airfoil coordinates) can enhance the predictive accuracy of the overall flow fields but cannot decrease the maximum error near the shock region. The addition of a filter in method D that enlarges the proportion of the flow fields near the airfoil surface decreases the minimum error by 40%, indicating the effectiveness of this approach.

2) The introduction of the gradient loss and multiwavelet transformation loss enhances the predictive accuracy, as these two loss functions capture the high-frequency component associated with the shock regions. Thus, the use of these loss functions decreases the average maximum error by nearly 50% and the maximum error near the shock wave from 0.04 to 0.02 or 0.01. The level of wavelet transformation must be appropriately set to balance the computational cost and loss benefits. In the considered scenarios, the second-order wavelet transformation loss achieves the highest performance with the lowest error.

3) Few-shot transfer learning (through fine-tuning) can achieve accuracy comparable with that obtained with conventional training but with considerably less computational time. When the model is pretrained on a tiny dataset, at least three snapshots must be used to achieve a precision of 4e-4. In contrast, models pretrained on small, medium, and large datasets require only one snapshot and achieve an accuracy of 1e-4. The model pretrained on the large dataset can predict the flow fields at a certain precision without fine-tuning. Therefore, the model can be readily applied in engineering scenarios.

Overall, the model can rapidly and accurately simulate the steady flow over supercritical airfoils. Future work can be aimed at enhancing the generalizability of the strategy to different grid topologies.

## Acknowledgments

## Appendix A

*A.I. Ablation experiments using various backbone models*

| Encoding method | Backbone model | Loss function | $l_{1\_avg}$ | $l_{1\_max\_avg}$ | $l_{1\_max\_max}$ |
|---|---|---|---|---|---|
| D | ViT | $l_1$ | 0.000187 | 0.0071 | 0.0687 |
| | U-net | $l_1$ | 0.000284 | 0.0133 | 0.1519 |

The ViT-based encoder–decoder outperforms the other backbone models.

*A.II. Ablation experiments using various loss functions*

| Encoding method | Loss function | $l_{1\_avg}$ | $l_{1\_max\_avg}$ | $l_{1\_max\_max}$ |
|---|---|---|---|---|
| D | $l_2$ | 0.000423 | 0.0142 | 0.0560 |
| | $l_1$ | 0.000187 | 0.00714 | 0.0687 |
| | $l_1 + \left|l_{grad}\right|_1$ | 0.000162 | 0.00384 | 0.0475 |

| | | | |
|---|---|---|---|
| $l_1 + \|l_{grad}\|_2$ | 0.000176 | 0.0038 | 0.0406 |
| $l_1 + \|l_{wavelet}\|_1$ | 0.000171 | 0.00430 | 0.0612 |
| $l_1 + \|l_{wavelet}\|_2$ | 0.000179 | 0.0054 | 0.0594 |

The results indicate the influence of different norms on the loss. The $l_1$-norm of the total loss is superior to the $l_2$-norm owing to the lower values of $l_{1\_avg}$ and $l_{1\_max\_avg}$. However, the losses combined with different norms of the gradient or wavelet loss are similar, possibly because both the losses assist the neural network to focus on the high-frequency components of the flow field, which comprise a small proportion of the complete flow field.

*A.III. Ablation experiments of different datasets*

| Encoding method | Dataset | | | Loss function | $l_{1\_avg}$ | $l_{1\_max\_avg}$ | $l_{1\_max\_max}$ |
| | Airfoils | AoA | Ma | | | | |
|---|---|---|---|---|---|---|---|
| D | 50 | 50–55 | - | $l_1 + l_{wt}$ | 0.000281 | 0.0138 | 0.1248 |
| | 500 | 50–55 | - | $l_1 + l_{wt}$ | 0.000171 | 0.00430 | 0.0612 |
| | 200 | 60 | 5 | $l_1 + l_{wt}$ | 0.000180 | 0.0063 | 0.124 |

As the number of airfoils increases, the accuracy of the model gradually increases. Additionally, the model can predict the flow fields when three different factors change simultaneously, which demonstrates its generalizability.